\documentclass{article}
\usepackage{float}
\usepackage[utf8]{inputenc}
\usepackage{authblk}
\usepackage{hyperref}
\usepackage{bm}
\usepackage{amsmath}
\usepackage{amsfonts}
\usepackage{amssymb}
\usepackage{array}
\usepackage{graphicx}

\title {Comment on "Comment on Galina Weinstein’s article entitled 'Convergences and Divergences: Einstein Poincaré and Special Relativity' "}
\author{Galina Weinstein\thanks{The Department of Philosophy, University of Haifa.}}

\begin{document}

\maketitle

\begin{abstract}
This paper provides a systematic response to the criticisms raised by Jean-Marc Ginoux in response to my review of his book on the history of relativity. 
Whereas my review was written in a strictly academic manner, Ginoux’s commentary intermingles mathematical objections with ad hominem insinuations about both Einstein and me. 
The purpose of this reply is twofold: first, to clarify the historiographical and conceptual issues at stake in the interpretation of Einstein’s 1905 paper and Poincaré’s contributions of 1905--1906; and second, to demonstrate why the conflation of algebraic form with conceptual content leads to distorted conclusions.  
\end{abstract}

\section{Introduction}

The present paper replies to Jean–Marc Ginoux’s extensive critique of my review of his book on the origins and reception of special relativity \cite{Gin}. Our exchange turns on a distinction that is as old as the subject but still too easily blurred. That is, the difference between \emph{formal structures}—equations, group properties, calculational devices—that were indeed available to Lorentz and Poincaré by mid–1905 and the \emph{conceptual framework} inaugurated by Einstein in June 1905, in which simultaneity is practically defined, the ether is rendered superfluous, and the Lorentz transformation is derived from two coequal postulates. My aim here is not to diminish Poincaré’s formidable contributions; it is to clarify what counts as founding a theory, and to separate reconstruction from documentation.

In what follows, I confine myself to dated publications, manuscripts, and verifiable correspondence. I avoid psychologizing—about Einstein, myself, or any other figure—and I do not treat private life or character judgments as evidence. Priority and influence should be argued from documents, not insinuations.

\section{Reply to Ginoux’s \emph{Ad Hominem} Critique} \label{ad-hominem}

In his comment, Ginoux challenges my interpretation of Einstein’s independence in 1905 and, in doing so, introduces remarks that go beyond scholarly disagreement. While my review \emph{was written in a strictly academic manner}, addressing only the content of his book, his reply includes \emph{ad hominem} statements about both Einstein and me. Such statements require clarification, as they risk obscuring the real historical issues through insinuations that are not only methodologically unsound but also derogatory.

Ginoux argues that Einstein’s retrospective statements—such as his 1955 letter to Carl Seelig, in which he wrote that he was unaware of Poincaré’s 1905 note \cite{Born}—should be dismissed because “Einstein, like many others, lied to his wife, his children, and also to his colleagues. So, why should we believe what he says about this article?” \cite{Gin}. This reasoning collapses private life into a wholesale claim of intellectual dishonesty. Methodologically, that is untenable; the personal failings of a scientist cannot be marshalled as evidence against their scholarly testimony. To reduce the question of influence in the genesis of special relativity to judgments about moral character is to leave the historical method for insinuation. As Einstein once remarked, “In the past, it never occurred to me that every casual remark of mine would be snatched up and recorded. Otherwise, I would have crept further into my shell” \cite{DH}.\footnote{I cite this remark only to explain my choice to avoid psychologizing and to keep the discussion on dated texts and derivations; it is not offered as proof of reliability. There is also a well-documented polemical tradition casting Einstein as dishonest or derivative. Without imputing such intent here, repeated emphasis on personal dishonesty risks resonating with that tradition. I therefore adhere to the textual record and refrain from personal judgments.}

The substantive historical debate is, of course, legitimate: Did Einstein, before mid-1905, have direct knowledge of Poincaré’s work on Lorentz invariance? Were his results anticipated, and if so, in what sense? On such questions, a substantial body of scholarship exists. Olivier Darrigol \cite{Dar}, Thibault Damour \cite{Dam}, Peter Galison \cite{Gal}, Gerald Holton \cite{Holton, Hol73}, Shaul Katzir \cite{Katzir}, Michel Janssen \cite{Jan}, Alberto Martínez \cite{Mar}, Arthur Miller \cite{Miller}, John Norton \cite{Nor}, Jürgen Renn \cite{Ren}, Renn $\&$ Hanoch Gutfreund \cite{GR}, and John Stachel \cite{Stachel2002, Stachel16}, among others, have thoroughly examined the textual record. As Holton has shown, Einstein’s 1905 derivation is conceptually distinct from both Lorentz and Poincaré, where the latter began with transformations to test for invariance; Einstein, on the other hand, deduced them from two postulates—the principle of relativity and the constancy of the speed of light. This difference is not incidental but foundational \cite{Holton}.

Ginoux dismisses Holton’s interpretation by suggesting that scholars often cite papers they have not read \cite{Gin}. While such a possibility exists in general, it does not invalidate Holton’s careful contextual analysis of Einstein’s citation practices across different works. The argument requires engagement with textual evidence, not dismissal by analogy. 

Disagreement over interpretation is a legitimate and welcome aspect of historiography. What is not acceptable is to dismiss Einstein’s retrospective testimony by means of personal denigration or to frame the debate in terms that echo long-standing prejudicial tropes. For the sake of scholarly integrity and historical accuracy, we must keep the focus where it belongs: on evidence, analysis, and conceptual interpretation.

\section{Against Ginoux’s "Fairy Tale" Claim} \label{formal}

Ginoux forcefully rejects the oft-repeated claim that Einstein stood outside the scholarly network until Planck “discovered” him after 1905. He calls this a “fairy tale” and “pure fiction.” According to Ginoux, the evidence suggests that Einstein was already well-integrated into the scientific community through \emph{Annalen der Physik}, where Planck served as associate editor, and where Einstein had been publishing since 1901. Ginoux writes that by 1905, Einstein had eight papers in that journal and was also reviewing articles for its supplement, the \emph{Beiblätter}.
Moreover, Einstein engaged directly with Planck’s work, reviewing a paper by Planck in 1906, and his correspondence clearly refers to letters from Planck before that year. In a letter dated 27 April 1906 to Maurice Solovine — at least if one follows Ginoux’s dating — Einstein explicitly states that Planck has recently written to him. This makes it impossible to claim that their first epistolary contact was as late as July 1907, as some historians suggest.
Ginoux concludes that Einstein and Planck were already in correspondence before April 1906, although those letters have since been lost. Thus, the narrative that Einstein was isolated from leading academics until his relativity paper drew attention is, in Ginoux’s formulation, nothing more than a “pure fiction” repeated from year to year by historians such as Galina Weinstein \cite{Gin}.

Ginoux misdates the Solovine letter: it is not from 27 April 1906, but from 6 May 1906. The relevant line reads \cite{ES}:
"Meine Arbeiten werden geschätzt und geben zu weiteren Untersuchungen Anla{\ss}. Herr Professor Planck (Berlin) hat mir k{\"u}rzlich dar{\"u}ber geschrieben."
(“My papers are appreciated and are giving rise to further investigations. Professor Planck (Berlin) has recently written to me about that.”)
This shows that Planck had contacted Einstein shortly before May 1906. 
It does not prove that their correspondence began before 1905. At most, it shows that Planck’s first letter reached Einstein sometime between 1905 and early 1906, which is perfectly consistent with the idea that Planck reacted to Einstein’s relativity paper after its publication. This supports a first contact no earlier than late 1905–early 1906, which is compatible with Planck's reaction to the printed 1905 papers. Ginoux therefore both errs in his citation of the letter and overstates its evidentiary weight.

Moreover, the claim that Einstein in 1905 lacked direct access to Poincaré’s May–June writings or to the inner correspondence networks of European physicists is not a "pure fiction" but the consensus of documentary historiography \cite{CPAE1, CPAE2, CPAE5}.
Historiography, like science, demands clarity. If a claim cannot be explained in simple, transparent terms, it is unlikely to rest on genuine understanding.

The \emph{Collected Papers of Albert Einstein} (CPAE) provides extensive coverage of Einstein’s early correspondence \cite{CPAE1, CPAE5}. There is no surviving letter from Lorentz to Einstein from this period, and Einstein’s first known exchange with Max Planck indeed dates back to 1906. 
Einstein's publications in the \emph{Annalen der Physik} before 1905 do not demonstrate integration into its editorial circles. Similarly, at the time of composing “On the Electrodynamics of Moving Bodies,” he was not part of the scholarly networks through which the Poincaré–Lorentz papers circulated. To state otherwise requires positive documentary evidence, not supposition.

\section{The Heuristic Method?}

Ginoux objects that I allow Einstein a heuristic derivation of the relativistic velocity–addition law while denying Poincaré a “heuristic” use of the ether, calling this a double standard: “It is surprising to read that Einstein has the right to use a heuristic method to obtain the \emph{relativistic velocity addition law} while Poincaré cannot use such a method to keep the \emph{luminiferous ether}” \cite{Gin}. But I never claimed anything so absurd. My review does not suggest that Poincaré was forbidden heuristic reasoning. For Einstein, “heuristic” designated a methodological device—a guiding principle for deriving new results within a framework that already dispensed with the ether. For Poincaré, by contrast, the ether was not a heuristic guide at all but an ontological commitment, a substantive entity posited as the medium of electromagnetic phenomena. To conflate these two uses of “heuristic” is to confuse method with substance.

\section{Who Stole the Lorentz Transformation?}

Ginoux raises two further points. First, he observes that Einstein only began referring to the transformation equations as “Lorentz transformations” around 1910, and speculates: “Starting from 1910, Einstein gave to the transformation he obtained in 1905 the name of Lorentz. … Why didn’t Einstein do it before? Probably another coincidence. The most probable reason is that if he had done that in 1905, it would have proven that he had read Poincaré” \cite{Gin}.

Here, Ginoux suggests that Einstein’s terminology was an act of concealment—that by not calling the formulas “Lorentz transformations” in 1905, he was hiding his dependence on Poincaré. Thus, Ginoux claims that by not calling the transformations “Lorentz,” Einstein deliberately avoided citing Poincaré. The record offers no such support. The irony is that the first person to publicly refer to them as “Lorentz transformations” was Poincaré himself, in his \emph{Comptes rendus} note of June 1905. Einstein submitted his paper on June 30, 1905. For Einstein to have deliberately avoided the term, he would have needed access to Poincaré’s unpublished draft—or advance notice from Poincaré, Lorentz, or the \emph{Comptes rendus} editorial board that Poincaré was about to use it. Short of imagining a private intelligence network devoted to suppressing credit, the timeline alone makes the suggestion untenable.

The more prosaic truth is that in 1905, Einstein called them “Koordinaten- und Zeittransformation” (“transformations of coordinates and time”) because his focus was on deriving them kinematically from his two postulates rather than embedding them in Lorentz’s electron theory. Only later did the label “Lorentz transformation” gain currency as the standard terminology, which Einstein naturally adopted thereafter. Naming conventions follow disciplinary consolidation; they are not smoking guns of concealed influence. To suggest otherwise is to confuse the history of terminology with the history of ideas.

\section{Same But Different}

Similarly, Ginoux objects: "The third type of argument is more dangerous but very classical: 'the falsehoods'. As an example, she explains that in his contribution entitled “La mesure du temps” Poincaré already used the concept of luminiferous ether although even the expression does not appear in this paper as it is easy to verify" \cite{Gin}.
At a deeper level, Ginoux attempts to reduce my analysis of Poincaré’s synchronization and Einstein’s synchronization to “same but different” \cite{Gin}.

Ginoux’s charge of “falsehood” rests on a narrow reading of "La mesure du temps." He notes, triumphantly, that the words “luminiferous ether” do not appear in that text. But absence of a phrase is not absence of a framework. In his 1900 Jubilee lecture for Lorentz, Poincaré’s synchronization by light signals—already sketched in 1898—was unambiguously set within Lorentz’s ether theory \cite{Poi00}. Local time was devised precisely to conceal motion relative to the ether. Thus, even when Poincaré interpreted local time pragmatically, it presupposed an underlying “true time” in the ether frame.

According to Galison, the question is not whether Einstein and Poincaré both employed signal-based synchronization—this was common practice in an era shaped by telegraphy and railroads—but how each conceptualized it \cite{Gal}:

\begin{quote}
Had Einstein seen Poincaré’s paper of 1898 or a crucial
subsequent one of 1900 before he wrote his 1905 paper? Possibly. While there is no definitive evidence one way or the other, it will, nonetheless, prove worthwhile to explore the question both narrowly and more widely. For as we will see, Einstein need not have read just those lines of Poincaré. Clock coordination appeared in the
pages of philosophy journals, and even occasionally in physics publications. 
\end{quote}

The issue is not the shared use of signal synchronization but its interpretation, nor whether Poincaré explicitly referred to the “ether” in 1898. In his “new mechanics,” Poincaré retained the ether as the substratum of “true time,” whereas Einstein declared it superfluous. That difference is neither semantic nor marginal, but the decisive reconfiguration that gave birth to special relativity.

\section{Einstein’s Method vs. Ginoux’s Historiography}

Ginoux objects that I ascribed to him the claim that his book adopts a formalist, sequence-oriented historiography, in which the systematic collation of equations, dates, and correspondence is used to reconstruct the relative timing and scope of contributions. On this basis, he attributes to Poincaré, by May–June 1905, a body of results that, combined with Poincaré’s articulation of the relativity principle, constitute the formal underpinnings of special relativity.
He insists instead that “this is not Ginoux who considered that this list of achievements constitutes the formal underpinnings of special relativity, but Einstein himself,” citing Einstein’s 1935 paper on mass–energy equivalence \cite{Gin}. 

In his paper "Elementary Derivation of the Equivalence of Mass and Energy," Einstein remarks \cite{Einstein35}: 

\begin{quote}
The question as to the independence of those relations is a natural one because the Lorentz transformation, the real basis of the special relativity theory, in itself has nothing to do with the Maxwell theory and because we do not know the extent to which the energy concepts of the Maxwell theory can be maintained in the face of the data of molecular physics.    
\end{quote}
 
Einstein’s point here is straightforward. Although the historical impetus for relativity grew out of electrodynamics (“the special theory of relativity grew out of the Maxwell electromagnetic equations”), the Lorentz transformations, once grounded in the relativity principle and the light postulate, stand as a purely kinematical structure, conceptually independent of Maxwell’s theory or Lorentz’s electron model. This independence is precisely why, in 1935, he can use nothing but the Lorentz transformations and the conservation laws of energy and momentum to derive $E = mc^2$ \cite{Einstein35}.

Ginoux, however, turns this methodological clarification into a historiographical claim. By treating Einstein’s statement as if it endorsed a purely formalist reconstruction of the origins of relativity, he aligns Einstein retrospectively with his own procedure of hoarding algebraic structures, sequencing equations, and collating correspondences. But Einstein’s remark was not a charter for such formalism. It was a methodological decision: to strip away electrodynamics and demonstrate that Lorentz invariance alone, coupled with conservation laws, is sufficient.
To read this as a historiographical endorsement is to misinterpret Einstein. His remark concerns the independence of a tool, not the historical path to a theory. Cleverness may lie in assembling formulas and chronological coincidences. But wisdom lies in recognizing when such parallels fail to capture the conceptual transformation that constitutes a genuine scientific revolution.

\section{The Title of Einstein's Paper}

Ginoux also highlights the verbal overlap between Poincaré’s 1904 Saint Louis lecture (quoting Poincaré: “It is a question before all of endeavoring to obtain a more satisfactory \emph{theory of the electrodynamics of moving bodies}”) and the title of Einstein’s 1905 paper, concluding: "Thus, is there any difference? Absolutely none" \cite{Gin}. 

But such coincidences of phrasing cannot be taken as evidence of transmission. Einstein himself acknowledged reading Poincaré’s \emph{Science and Hypothesis} \cite{Poi02} before 1905. But there is no documentary evidence that he ever consulted the 1904 Saint Louis lecture, nor the sentence on page 319: "Ne devrions-nous pas aussi nous efforcer d’obtenir une théorie plus satisfaisante de l’électrodynamique des corps en mouvement?" (“Should we not also strive to obtain a more satisfactory theory of the electrodynamics of moving bodies?” \cite{Poi04}) There are no italics on "l'électrodynamique des corps en mouvement" in the original printing. More importantly, the resemblance lies only in phrasing. The substantive difference is in the architecture of the papers: Poincaré framed electrodynamics within an ether-bound dynamics, whereas Einstein reframed it as a principle-based kinematics, dispensing with the ether altogether.

\section{Heuristic Freedom, Not Foreknowledge}

Ginoux argues that Einstein’s introduction of the factor:

\begin{equation} \label{pa}
\varphi(v) = a(v)\,\gamma,    
\end{equation}
was not a neutral reparametrization but a purposeful nudge toward the Lorentz transformation, echoing Arthur I. Miller’s remark that it seems as if Einstein knew beforehand the correct form of the set of relativistic transformations \cite{Miller, Gin}. 

These criticisms warrant clarification.
The redefinition \eqref{pa} is best understood as a mathematical convenience, not as a presupposition of the final result. 
Einstein’s derivation proceeds step by step: from the synchronization condition and the light postulate, to the linear form with an undetermined scale factor $a(v)$, then to the consolidated form with $\varphi(v)$, and only finally to the determination $\phi(v)=1$ by reciprocity, isotropy, and continuity at $v=0$. 
This sequence --- synchronization condition $\Rightarrow$ partial differential equation $\Rightarrow$ general solution $\Rightarrow$ linearity $\Rightarrow$ provisional transformation $\varphi(v)$  $\Rightarrow$ $\varphi(v)=1$ --- shows that Einstein did not begin with the Lorentz transformation in hand, but allowed the equations to retain an undetermined freedom until physical constraints eliminated it. 
The introduction of $\varphi(v)$ isolates the single surviving degree of freedom so that the relativity principle can cleanly fix it. 
It is not evidence of "foreknowledge," but precisely the opposite: the carrying through of an arbitrary function until physics forces it to unity. 

\section{On the Relativity of Irony}

Ginoux writes: “First of all, it is not only Ginoux who ‘treats \eqref{pa} as ...’ but also Professor Arthur I. Miller. Then, it is unclear how and why Einstein introduced this factor as recalled by Miller who wrote in 1981” \cite{Gin}.

Here, the irony could hardly be more delicate. Miller—the very historian whose authority Ginoux invokes— places, in the introduction to Ginoux’s own book, the remark \cite{Gin1}:

\begin{quote}
The problem can be viewed in three different ways: focusing on the scientific papers of the two men 
around 1905, in order to disentangle their scientific procedures; investigating the ways in which their views of space and time differed; and analysing why Einstein 
discovered the special theory of relativity and not Poincaré even though they both used the same mathematical equations.    
\end{quote}

The passage sits in plain view, quietly inverting the very argument Ginoux has been eager to defend. We almost hear Einstein saying, "Subtle is the irony, but vicious it is not."

\section{A Long and Tedious Computation}

Ginoux further claims that my reading of Einstein’s derivation of the relativistic addition law is mistaken, writing \cite{Gin}:
\begin{quote}
At the end of this subsection, Galina Weinstein makes a new mistake by writing: “Using the final Lorentz transformation (38), after fixing $a(v)$ and $\phi(v)=1$, he obtained the relativistic addition law \cite{Einstein05}.” Contrary to what Galina Weinstein has written, in his original paper, Einstein first used a long and tedious computation to obtain the relativistic addition law.
\end{quote}

Ginoux reinforces his objection by reproducing a screenshot of a page from Einstein’s derivation in section §5, and captioning it: “As one can see in Fig. 1, contrary to what claims Galina Weinstein, this is not by using the final Lorentz transformation that Einstein first obtained the relativistic addition law”  \cite{Gin}. 

This presentation, unfortunately, rests on a mistaken premise of the mathematics at issue. What Ginoux calls a “long and tedious computation” is nothing more than the straightforward working-out of velocity components from those equations.
Ginoux’s insistence that Einstein’s derivation in section §5 of the velocity addition law is somehow not based on the Lorentz transformation rests on a fundamental misunderstanding. The calculation in section §5 is precisely Einstein’s way of applying the transformation—already fixed with $\phi(v)=1$, to the motion of a particle in system $k$. By transforming a straight-line trajectory $\xi = w_\xi \tau$, $\eta = w_\eta \tau$ into the $K$ system, and then dividing space by time coordinates, Einstein arrives directly at the velocity transformation laws.
Einstein's steps in section \S 5 are transparent (see my reconstruction in \cite{Wein}): start with the inverse Lorentz transformation, substitute the particle’s motion, and divide to obtain the components. The algebra is straightforward once $\phi(v)=1$ is imposed. What Ginoux calls a “long and tedious computation” is in fact only the routine differentiation and substitution that any student of relativity would work through on the blackboard.

Einstein’s genius here lies not in tedious manipulation. Still, in the heuristic method, he carried an arbitrary factor through the transformation, fixed it by reciprocity and isotropy, and then let the physics itself deliver the velocity law. This was not foreknowledge but disciplined restraint—a refusal to assume more than his postulates warranted until the argument compelled it.

Thus, the derivation in section §5 is not evidence of some hidden dependence or a “mistake” in my reading. It is precisely what it looks like: Einstein, having secured the Lorentz transformation in final form, applies it to particle motion and uncovers the relativistic law of velocity composition. On this point, the mathematics leaves no room for doubt.
As Einstein himself might have reminded us, the equations are subtle, but not malicious.

\section{Einstein's Citation Practice}

Ginoux insists that my use of Einstein’s reply to Seelig—stating that before 1905 he knew only Lorentz’s \emph{Versuch} \cite{Lorentz1895} and not Poincaré’s papers \cite{Born}—is misleading. He argues that Einstein’s March 1905 light-quantum paper contained numerous references, which shows that Einstein was well-versed in citation practices and therefore could (and should) have cited Poincaré and Lorentz's 1904 paper \cite{Lorentz1904} in June as well: “I still claim that it’s very surprising that Einstein ignored the work of his predecessors in this article entitled ‘On the Electrodynamics of Moving Bodies,’ while in his seven other articles published between 1901 and 1905 in \emph{Annalen der Physik} he cited all the most important work of his predecessors. … He was therefore well-versed in the scientific publication practices of his day. … So why does he quote his peers in March and not in June? Once again, this fact raises questions” \cite{Gin}.

This mode of argumentation does not advance historical understanding; it is an inference that does not follow from the evidence. The absence of citations in Einstein’s June paper is a well-known fact, and historians have long debated its significance. But to frame the issue as a “raised question” without evidence is to insinuate concealment rather than to demonstrate it. The difference in citation practice between Einstein’s March and June 1905 papers reflects their differing aims and genres: the former was embedded within contemporary experimental and theoretical literature on radiation, while the latter built a principle-based kinematics whose originality lay precisely in bracketing the electrodynamical tradition. To suggest that Einstein’s silence was a strategy of suppression is not historiography but innuendo. The historian’s task is not to treat absence as guilt but to test claims against the surviving documentary record.

Galison has offered a subtle explanation for Einstein’s style in the 1905 relativity paper. Einstein had been trained in the patent office, where clarity, compression, and originality were paramount. Patent applications never cite prior patents or scientific works, for the very logic of the system demands that the invention stand on its own, free of genealogical entanglements. As Galison observes, in the fifty or so Swiss electric clock patents filed around 1905, not one contains a single footnote. Immersed for three years in this culture under Friedrich Haller’s exacting standards, Einstein absorbed the “precise and austere” patent style as his natural mode of exposition.
Seen in this light, the absence of citations in “On the Electrodynamics of Moving Bodies” is not evidence of concealment, nor proof of surreptitious borrowings, but a reflection of the genre Einstein had internalized \cite{Gal}. Unlike the relativity paper, with its underlying resonance with clocks and patents, the March 1905 light-quantum paper was not framed by such practical imagery. Though heuristic, it still bore the marks of a physics article replete with references, contextual framing, and engagement with contemporary debates.

\section{Einstein’s 1920 "Ether" Was Not Lorentz’s Ether}

Ginoux contends that Einstein did not abolish the ether in 1905, but merely declared it "superfluous," and that his later remarks in the 1920 Leiden lecture and in an unpublished article for \emph{Nature} show that the ether was reintroduced into general relativity \cite{Gin}. This line of argument conflates radically different uses of the term "ether" and overlooks the decisive conceptual break that Einstein himself insisted upon.

Ginoux’s argument hinges on Einstein’s use of the word überflüssig (“superfluous”) in his 1905 relativity paper, as if this meant that Einstein only sidestepped the ether rather than abolishing it \cite{Gin}. This reading is misleading. In German scientific usage, überflüssig means not “almost unnecessary,” but “dispensable,” “without function,” “obsolete.” When Einstein writes that the introduction of a “light ether” will prove überflüssig, he is saying that the ether is not required to account for electrodynamic phenomena. That is abolition in the strictest physical sense.
To suggest otherwise, as Ginoux does, is to flatten the conceptual break between Einstein and Poincaré. For Poincaré, the ether was indispensable because it carried the stresses that preserved action–reaction and grounded “true time” beneath conventions of synchronization. For Einstein, simultaneity itself became a constitutive definition, with no underlying substrate. That is the point of calling the ether superfluous; it has no remaining role in the theory.

Einstein was not an ether enthusiast in hiding, pretending to renounce what he secretly believed in. His program was kinematical, not dynamical. By contrast, Poincaré openly maintained the ether precisely because his program was constructive and dynamical. To turn Einstein’s methodological clarification into a charge of disingenuousness is to confuse rhetoric with physics.

The real point of contention is not whether Poincaré made profound contributions—he did—but whether Einstein’s 1905 paper represents merely an echo of those contributions or a decisive reconceptualization. I maintain, in line with much of the existing scholarship (\cite{Nor, Stachel2002} and others), that Einstein’s achievement lay in relocating the Lorentz transformations into an ether-free, principle-based kinematics. This constitutes not just a change of form but a transformation of the theoretical architecture.

In his Leiden inaugural lecture, Einstein spoke of an ether in a sense that differed fundamentally from Lorentz’s construct \cite{CPAE9}, Doc. 373. Einstein described a “new ether” or “world-matter,” a Machian medium required as a carrier of inertial effects in a universe where action at a distance was no longer admissible. This was not Lorentz’s rigid, absolute frame of reference. Einstein referred to the metric field $g_{\mu\nu}$ itself—varying from place to place, determined by material phenomena, and never assignable to an independent state of motion.

Einstein’s remark in his letter to Lorentz in 1916 that general relativity was "closer to the ether hypothesis" than special relativity \cite{CPAE8}, Doc. 222, and his rhetorical flourish in Leiden about the ether, must be read in their historical context. Lorentz still clung to the ether, and Einstein, who revered him, chose to frame his redefinition of the \emph{metric field} of general relativity in language Lorentz could recognize. As Stachel has noted, the ether that Einstein reintroduced "differed fundamentally from the ether he had banished" \cite{Stachel2001}. To ignore this discontinuity is to replace conceptual analysis with a slogan.

\section{Einstein’s Light Postulate: An Indispensable Principle, Not a Curiosity}

Ginoux reiterates his view that Einstein’s 1905 light postulate is “curious,” since in his opinion the invariance of $c$ ought to have followed directly from the relativity principle, as in Poincaré’s formulations : "I still claim that Einstein’s 1905 light postulate is curious, since in my opinion it should have been like for Poincaré a consequence of the \emph{relativity principle} and not the contrary" \cite{Gin}.
He then invokes Poincaré’s 1904 Saint Louis lecture \cite{Poi04} (“no velocity could surpass that of light”) and the choice of units $c=1$ in Poincaré’s 1905 correspondence with Lorentz as evidence that the invariance of $c$ was already implicit in Poincaré’s writings: “Obviously, if Poincaré posed $c=1$, it is because he considered the velocity of light was the same in all reference frames” \cite{Gin}.

Ginoux’s objection collapses under its own weight. He criticizes Einstein for formulating the constancy of $c$ as an independent postulate, insisting that it should have been, as in Poincaré’s writings, a consequence of the relativity principle. Yet the very passages he cites as evidence—from the Saint Louis lecture and from Poincaré’s correspondence with Lorentz—reveal the opposite.%
\footnote{Poincaré states on page 313 of the Saint Louis lecture: "... leur masse croît avec la vitesse, en sorte qu’il n’est plus possible de leur faire dépasser la vitesse de la lumière” (“… their mass increases with velocity, so that it is no longer possible to make them exceed the velocity of light”) \cite{Poi04}.}
The bound $v<c$ in 1904 was a dynamical result of Lorentz’s electron theory, not a kinematical deduction from relativity alone. Likewise, the adoption of $c=1$ was a notational convention, not a proof. Einstein’s great leap was precisely to elevate the invariance of $c$ to a principle coordinate with relativity itself, thereby disentangling it from Lorentzian dynamics and making it constitutive of a new kinematics. Far from being “curious,” this move marked the conceptual break that Ginoux’s reading obscures.

\section{Algebraic Form Is Not Physical Content}

Ginoux argues that my position is contradictory: "This is simply incredible! The first argument of Galina Weinstein ... was that Poincaré used group symbols. But here, $\xi$ and $\varepsilon$ are still dimensionless group symbols. So, according to her own argument":

\begin{equation} \label{xi}
\xi' = \frac{\xi + \varepsilon}{1+\varepsilon \xi}
\end{equation}
"should not be considered as the relativistic velocity addition law. Could Galina Weinstein explain us the difference between" \eqref{xi} and:

\begin{equation} \label{varep}
\varepsilon''=\frac{\varepsilon+\varepsilon'}{1+\varepsilon\varepsilon'},    
\end{equation}
above" \cite{Gin}.

Ginoux’s remarks in this passage rely on rhetorical questions and exclamations ("This is simply incredible!") rather than substantive analysis. I do not consider it appropriate to respond to arguments framed in such derogatory language. For a careful treatment of the velocity addition law in both Einstein’s and Poincaré’s work, the reader may consult my review \cite{Weinstein}.

\section{On the Alleged "Coincidence" with Lorentz and Poincaré}

Ginoux contends that Einstein could not have known the correctness of his section \S 9 expressions for charge density and convection current without consulting Lorentz’s 1904 paper \cite{Lorentz1904}, whose source formulas he deems “erroneous” and “later corrected by Poincaré." His case rests on two assertions \cite{Gin}:
\begin{enumerate}
  \item Lorentz’s 1904 expressions for velocity and charge density were wrong; and:
  \item Einstein’s results are “the same as those obtained in May--June 1905 by Poincaré,” hence dependence. 
\end{enumerate}
I address (1) and then (2).

\medskip

(1) Ginoux labels Lorentz’s 1904 formulas for charge density and velocity as “erroneous expressions,” to be “corrected” by Poincaré. That framing commits a category mistake. In a letter of May 1905, Poincaré writes to Lorentz that, while he agrees on the essentials, there are “\emph{quelques divergences de d\'etail}.”\footnote{“a few differences of detail”.} He proceeds to propose \emph{replacements} for Lorentz’s page 813 relations for $\rho'$ and the $x$-component of the current/velocity \cite[letter 38.3]{Wal}. The phrasing is explicit: “\emph{au lieu de poser} ... \emph{il me semble qu'on doit poser} ...”\footnote{“instead of setting ... it seems to me that one ought to set ...}; the stated motivation is to secure conservation of the (apparent) electron charge, \emph{si l'on veut que la charge apparente de l'\'electron se conserve}”\footnote{“if one wants the apparent charge of the electron to be conserved”.} with $\varepsilon=-w/c$ (or $\varepsilon=-w$ for $c=1$) \cite{Wal,Lorentz1904}. 

In his \emph{Comptes Rendus} note (5 June 1905) Poincaré credits Lorentz and says he largely agrees \cite{Poi05-1}:%
\footnote{“The importance of the question led me to take it up again; my results agree with Lorentz’s on all essential points; I have only been led to modify and complete them in a few details.”}

\begin{quote}
L'importance de la question m’a déterminé à la reprendre; les résultats que j'ai obtenus sont d’accord sur tous les points importants avec ceux de Lorentz; j'ai été seulement conduit à les modifier et à les compléter dans quelques points de détail.  
\end{quote}

Poincaré nowhere characterizes Lorentz’s source transformations as “erroneous.” In both the May 1905 letter and the \emph{Comptes Rendus} note, he speaks of \emph{modifying} and \emph{completing} Lorentz’s results “in a few details,” while explicitly naming the transformation “Lorentz” and agreeing on the essentials. 

According to Michel Janssen, for \textit{Lorentz} \cite{Lorentz1904}, the primed symbols:
\begin{equation} \label{prime}
x',t',\rho',\mathbf \rho' u',    
\end{equation}
are auxiliary variables in the theorem of corresponding states introduced to keep Maxwell’s equations formally invariant; they are not what a co-moving observer’s instruments would actually read. 
For \textit{Einstein} \cite{Einstein05}, the primed quantities \eqref{prime} denote actual measurements in an inertial frame $k$ moving with speed $v$ relative to frame $K$. 
For \textit{Poincaré} \cite{Poi00, Poi05}, the primed quantities \eqref{prime} denote magnitudes in the system moving with velocity $v$ relative to the ether \cite{Jan}. 

Already in 1900 \cite{Poi00}, Poincaré read Lorentz’s "local time" as the time indicated by clocks of an observer moving with respect to the ether (with a background "true" ether time still assumed). By the same token, Janssen argues that in 1905 \cite{Poi05-1}, Poincaré \emph{reinterpreted} Lorentz’s 1904 primed symbols \eqref{prime} as quantities measured in the moving system, and from this vantage point, he was "led to modify and complete them in a few details." But within Lorentz's 1904 theorem of corresponding states, the primed quantities \eqref{prime} made perfect sense and were not wrong at all \cite{Jan}.

(2) Ginoux's objection dissolves once one actually follows Einstein’s 1905 derivations (see the derivations in my paper \cite{Weinstein}). In section \S9, Einstein deals with inhomogeneous Maxwell–Hertz equations with sources. By transforming the differential operators, applying the field transformations, and invoking the continuity equation, he arrives at the charge–density and current transformations \cite{Einstein05}:
\begin{equation} \label{TC}
\rho' = \gamma \left(\rho - \frac{v}{c^{2}}\rho u_x\right), \quad 
\rho'u'_{\xi} = \gamma(\rho u_x - v\rho), \quad
\rho'u'_{\eta} = \rho u_y, \quad
\rho'u'_{\zeta} = \rho u_z.
\end{equation}
Dividing the components of current by $\rho'$ then reproduces the velocity transformations \cite{Einstein05}:
\begin{equation}
u'_{\xi}= \frac{u_x - v}{1 - \tfrac{u_x v}{c^2}}, \quad  
u'_{\eta} = \frac{u_y}{\gamma \left( 1 - \tfrac{u_x v}{c^2} \right)}, \quad
u'_{\zeta} = \frac{u_z}{\gamma \left( 1 - \tfrac{u_x v}{c^2} \right)}.
\end{equation}
On this basis, Einstein could certify that his results were correct, independently of Lorentz \cite{Lorentz1904} or Poincaré’s letters and publications.  

The crucial point is methodological. Poincaré did not arrive at the velocity transformations by dividing the transformed currents \eqref{TC} by $\rho'$. This step, simple but decisive, appears first in Einstein’s treatment of the continuity equation and Ampère–Maxwell law \cite{Einstein05}. In his \emph{Rendiconti} memoir, Poincaré did obtain the correct velocity transformation formulas. Still, by a different route, he began from the Lorentz transformations themselves and formed the ratios $\frac{dx'}{dt'}$, $\frac{dy'}{dt'}$, $\frac{dz'}{dt'}$ \cite{Poi05}. Thus, while the algebraic structures were within reach in May 1905, Poincaré did not execute Einstein’s particular derivation. What matters historiographically is not the latent algebra but the explicit reasoning carried out in print. 

\section{On the Originality of My Reconstructions}

Ginoux asserts that all the mathematical derivations I present in \cite{Weinstein} were already contained in his book and in Miller’s works: "Unfortunately, all the mathematical derivations she pretends to 'reconstruct' ... have been already presented in my book in more detailed way at Chapter 6 ... and in Miller’s contributions. Thus, her 'novel way' dates back at least of 1973 and may be before" and: "First of all she explains that she will “reconstruct in a novel way the 1905 derivations of Einstein and Poincaré” although she had only recopied some results already published in my book and in the contributions of Pr. Miller [17–20]. Thus, her 'novel way' dates back at least of 1973 and may be before \cite{Gin}.

Ginoux writes that "she had only recopied some results already published in my book and in the contributions of Pr. Miller." Let me be absolutely clear: this is not only false, it is a serious insinuation of plagiarism. To suggest that my work consists of "recopying" is an attack on my scholarly integrity.

What I actually present in my paper \cite{Weinstein} is a systematic reconstruction of Einstein’s derivational pathway from the Lorentz transformation through to the velocity transformations and source terms. This scaffolding \emph{is not} to be found in Miller’s works, \emph{nor} in Ginoux’s. Their contributions contain formulas and derivations, but not the coherent, step-by-step logical architecture I provide. The distinction between scattered formulas and a full reconstruction is fundamental.

To equate my work with mere "recopying" is therefore doubly mistaken:

\begin{enumerate}
    \item It misrepresents the substance of my contribution, which is the original organization, scaffolding, and interpretive framework of the derivations.
    \item It impugns my integrity by implying that I appropriated material without originality. Such an insinuation is not only unfounded but unacceptable in scholarly discourse.
\end{enumerate}

I emphasize that originality does not reside in the presence of individual formulas, which can be found across many sources. It resides in the logic, order, and reconstruction—in the intellectual architecture that allows us to understand Einstein’s reasoning afresh. That is what my work contributes, and that is what has not been done before.

If Ginoux believes otherwise, the burden is on him to demonstrate, with precision and line-by-line comparison, where in Miller or in his own book my full derivational scaffolding already appears. Until he does so, his claim of "recopying" is not scholarship; it is polemics.

As Einstein remarked in a letter to Marcel Grossmann, “God created the donkey and gave him a thick skin” (“Gott schuf den Esel und gab ihm ein dickes Fell”) \cite{CPAE1}, Doc. 100.%
\footnote{I quote it only as a stylistic aside about scholarly thick skin, not as evidence on priority.}
I will continue to meet polemical attacks with patience, but also with unwavering commitment to rigor and integrity.

\section{Historiography Is Not Rhetorical Characterization}

Ginoux concludes his criticism not with evidence but with rhetoric, attempting to delegitimize my work by categorizing it as “inventing imaginary facts” or “fake news.” This is not historical argumentation; it is polemic. Historiography is not advanced by branding one’s interlocutors as “regrettable,” but by engaging with sources and reasoning. 

Ironically, Ginoux closes his article by invoking Poincaré’s maxim that says that thought must submit to facts alone \cite{Poi09}:

\begin{quote}
La pensée ne doit jamais se soumettre, ni à un dogme, ni à un parti, ni à une passion, ni à un intérêt, ni à une idée préconçue, ni à quoi que ce soit, si ce n’est aux faits eux-mêmes, parce que, pour elle, se soumettre, ce serait cesser d’être.    
\end{quote}
\begin{quote}
Thought must never submit itself, neither to a dogma, nor to a party, nor to a passion, nor to an interest, nor to a preconceived idea, nor to anything whatsoever, except to the facts themselves; for to submit would be to cease to exist.    
\end{quote}

That principle is precisely what I have upheld throughout my analysis: adherence to the extant documentary record, refusal to speculate on lost or hypothetical letters, and attention to the conceptual architecture of theories rather than to coincidences of terminology or algebra. To insist, without evidence, that Einstein concealed knowledge of Poincaré, or that citation practices prove suppression, is not to “submit to facts,” but to subordinate facts to suspicion. 

Einstein himself once remarked that he had “no special talent, only passionate curiosity” \cite{Seelig}. His independence in 1905 lay not in inherited knowledge but in that very curiosity, carried step by step through to its logical conclusions. If one wishes to debate that independence, the appropriate arena is the archive, not the rhetoric of dogma and denunciation. Scholarly integrity demands as much. 

\section{A Final Word, and It Is Einstein’s}

Finally, Ginoux writes: "Then, Galina Weinstein explains that:  .... Poincaré’s own words, in the last public address of his life, sound less like a claimant and more like a convert. ... This is again false. Poincaré gave his last lecture at the École supérieure des Postes et Télégraphes today SupTelecom Paris in July 1912, a few days before his death, as confirmed by the subtitle" \cite{Gin}. 

Ironically, as Poincaré’s own lecture shows (see Fig. \ref{FP}), his final public remarks—delivered at the threshold of death at the École supérieure des Postes et Télégraphes in July 1912—explicitly mention Einstein and were published contemporaneously \cite{Poi12}:

\begin{figure}[ht] 
    \centering
    \includegraphics[width=0.8\textwidth]{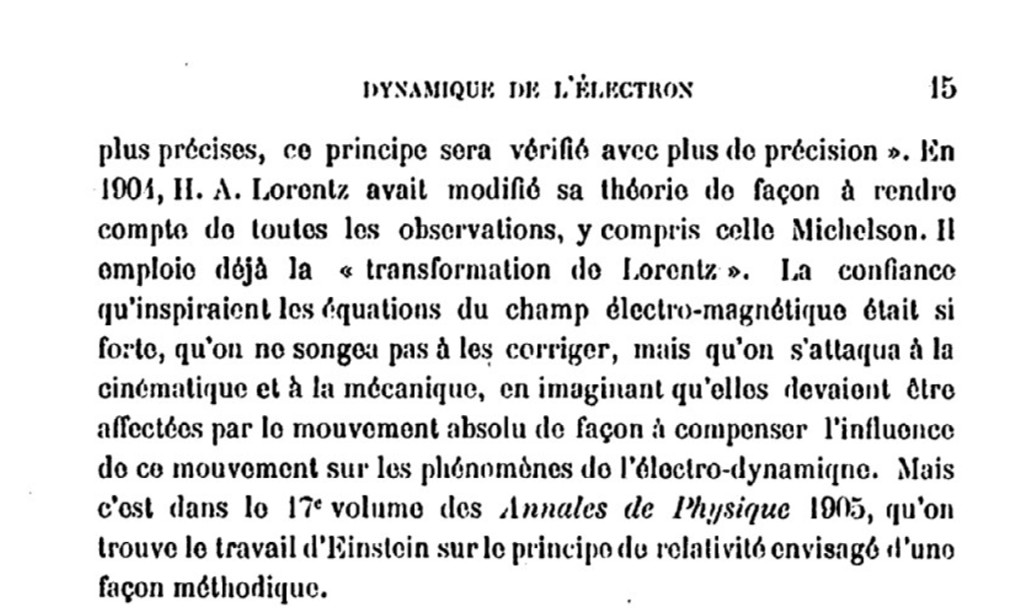} 
    \caption{Excerpt from Poincaré’s 1912 lecture at the École Supérieure des Postes et Télégraphes.}
    \label{FP}
\end{figure}


\newpage

\begin{thebibliography}{18}

\bibitem[Bor]{Born} Born, M. (1957) \emph{Physik im Wandel meiner Zeit}. Braunschweig: Vieweg.

\bibitem[CPAE1]{CPAE1} \emph{Collected Papers of Albert Einstein. Vol. 1: The Early Years,1879–1902}. Stachel, J., Cassidy, D. C., and Schulmann, R. (eds.). Princeton: Princeton University Press, 1987.  

\bibitem[CPAE2]{CPAE2} \emph{The Collected Papers of Albert Einstein. Vol. 2: The Swiss Years: Writings, 1900–1909}. Stachel, J., Cassidy, D. C., and Schulmann, R. (eds.). Princeton: Princeton University Press, 1989. 

\bibitem[CPAE5]{CPAE5} \emph{Collected Papers of Albert Einstein. Vol. 5: The Swiss Years: Correspondence, 1902–1914}. Klein, M. J., Kox, A. J., and Schulmann, R. (eds.). Princeton: Princeton University Press, 1993. 

\bibitem[CPAE8]{CPAE8} \emph{Collected Papers of Albert Einstein. Vol. 8: The Berlin Years: Correspondence, 1914–1918}. Schulmann, R., Kox, A.J., Janssen, M., Illy, J. (eds.). Princeton: Princeton University Press, 2002.

\bibitem[CPAE9]{CPAE9} \emph{Collected Papers of Albert Einstein. Vol. 9: The Berlin Years: Correspondence, January 1919–April 1920}. Buchwald Kormos, D., Schulmann, R., Illy, J., Kennefick, D. J., and Sauer, T. (eds.). Princeton: Princeton University Press, 2004.

\bibitem[Dam]{Dam} Damour, T. (2017). "Poincaré, the dynamics of the electron, and relativity." \emph{Comptes Rendus Physique} 18, pp. 551–-562.

\bibitem[Dar]{Dar} Darrigol, O. (2022). \emph{Relativity Principles and Theories from Galileo to Einstein}. Oxford: Oxford University Press.

\bibitem[DuHo]{DH} Dukas, H. and Hoffmann, B. (1979). \emph{Albert Einstein, the Human Side: New Glimpses from his Archives}. Princeton: Princeton University Press. 

\bibitem[Ein05]{Einstein05} Einstein, A. (1905). "Zur Elektrodynamik bewegter Körper." \emph{Annalen der Physik} 17, pp. 891--921.

\bibitem[Ein35]{Einstein35} Einstein, A. (1935). "Elementary Derivation of the Equivalence of Mass and Energy." \emph{Bulletin of the American Mathematical Society} 37, pp. 39–-44.

\bibitem[EinSol]{ES} Einstein, A. and Solovine, M. (1956). \emph{Lettres à Maurice Solovine}. Paris: Gauthier-Villars.

\bibitem[Gal]{Gal} Galison, P. (2003). \emph{Einstein's Clocks, Poincaré's Maps. Empire of Time}. New York: W. W. Norton $\&$ Co. 

\bibitem[Gin-1]{Gin1} Ginoux, J-M (2024). \emph{Poincaré, Einstein and the Discovery of Special 
Relativity. An End to the Controversy.} Cham: Springer.  

\bibitem[Gin-2]{Gin} Ginoux, J-M (2025). "Comment on Galina Weinstein’s article entitled “Convergences and Divergences: Einstein Poincaré and Special Relativity”." \emph{ArXiv}2509.22726v1  [physics.hist-ph], pp. 1--14.  

\bibitem [GutRe] {GR} Gutfreund, H. and Renn, J. (2023).
\emph{The Einsteinian Revolution: The Historical Roots of His Breakthroughs}. Princeton: Princeton University.

\bibitem[Hol60]{Holton} Holton, G. (1960). "On the Origins of the Special Theory of Relativity." \emph{American Journal of Physics} 28, pp. 627–-636.

\bibitem[Hol73]{Hol73} Holton, G. (1973). \emph{Thematic Origins of Scientific Thought Kepler to Einstein}. Cambridge, MA: Harvard University Press. 

\bibitem[Kat]{Katzir} Katzir, S. (2005). "Poincaré’s Relativistic Physics: Its Origins and Nature." \emph{Physics in Perspective} 7, pp. 268–-292.

\bibitem[Jan]{Jan} Janssen, M. (2019). "How did Lorentz Find his Theorem of Corresponding States?" \emph{Studies in History and Philosophy of Modern Physics} 67, pp. 167–-175. 

\bibitem[Lor85]{Lorentz1895}
Lorentz, H. A. (1895). \textit{Versuch einer Theorie der elektrischen und optischen Erscheinungen in bewegten Körpern}. Leiden: E. J. Brill.

\bibitem[Lor04]{Lorentz1904}
Lorentz, H. A. (1904). "Electromagnetic Phenomena in a System Moving with any Velocity Smaller than that of Light." \emph{Verslagen van de Koninklijke Akademie van Wetenschappen te Amsterdam, Sectie Natuurkunde} 6, pp. 809--836.

\bibitem[Mar]{Mar} Martínez, A. A. (2009). \emph{Kinematics. The Lost Origins of Einstein’s Relativity.} Baltimore: Johns Hopkins University Press. 

\bibitem[Mil]{Miller} Miller, A. I. (1981). \emph{Albert Einstein’s Special Theory of Relativity: Emergence (1905) and Early Interpretation (1905–1911)}. Reading, Massachusetts: Addison-Wesley.

\bibitem[Nor]{Nor} Norton, J. (2004). "Einstein's Investigations of Galilean Covariant Electrodynamics prior to 1905." \emph{Archive for the History of Exact Sciences} 59, pp. 45--105.

\bibitem[Poi98]{Poi98} Poincaré, H. (1898). "La mesure du temps." \emph{Revue de métaphysique et de morale} 6, pp. 371--384.

\bibitem[Poi00]{Poi00} Poincaré, H. (1900). "La théorie de Lorentz et le principe de réaction." \emph{Archives néerlandaises des sciences exactes et naturelles} 5, pp. 252--278.

\bibitem[Poi02]{Poi02} Poincaré, H. (1902). \emph{La science et l’hypothèse}. Paris: Ernest Flammarion.

\bibitem[Poi04]{Poi04} Poincaré, H. (1904). "L'état actuel et l'avenir de la physique mathématique." \emph{Bulletin des Sciences Mathématiques} 28, pp. 302---324.

\bibitem[Poi05-1]{Poi05-1} Poincaré, H. (1905). "Sur la dynamique de l'électron." \emph{Comptes rendus des séances de l'Académie des sciences} 140, pp. 1504--1508. 

\bibitem[Poi05-2]{Poi05} Poincaré, H. (1906). "Sur la dynamique de l'électron." \emph{Rendiconti del Circolo Matematico di Palermo} 21, pp. 1--47. 

\bibitem[Poi09]{Poi09} Poincaré, H. (1909). "Le libre examen en matière scientifique." \emph{Revue de l’Université de Bruxelles} 15, pp. 285–-295.

\bibitem[Poi12]{Poi12} Poincaré, H. (1912). "La dynamique de l'électron." \emph{Bibliothèque des Annales des postes, télégraphes et téléphones}. Paris: Dumas.

\bibitem[Renn]{Ren} Renn, J. (1993). "Einstein as a Disciple of Galileo. A Comparative Study of Concept Development in Physics." \emph{Science in Context} 6, pp. 311--341.

\bibitem[See54]{Seelig} Seelig, C. (1954). \textit{Albert Einstein. Eine dokumentarische Biographie}. Zurich: Europa Verlag.

\bibitem[Sta-01]{Stachel2001} Stachel, J. (2001). "Why Einstein Reinvented the Ether." \emph{Physics World} 14, pp. 55--56.

\bibitem[Sta02]{Stachel2002} Stachel, J. (2002). \textit{Einstein from 'B' to 'Z'}. Washington D.C.: Birkhäuser.

\bibitem[Sta16]{Stachel16} Stachel, J. (2016). "Poincaré and the Origins of Special Relativity." \emph{HOPOS: The Journal of the International Society for the History of Philosophy of Science} 6, pp. 242--256.

\bibitem[WBC]{Wal} Walter, S., Bolmont, É. et Coret, A. (2000). \emph{La correspondance entre Henri Poincaré et les physiciens, chimistes et ingénieurs}. Berlin: Birkhäuser.

\bibitem[Wei25-1]{Wein} Weinstein, G. (2025). "Einstein’s Hidden Scaffolding, with a Glance at Poincaré." \emph{ArXiv}:2509.02456v1  [physics.hist-ph], pp. 1--34.

\bibitem[Wei25-2]{Weinstein} Weinstein, G. (2025). "Convergences and Divergences: Einstein, Poincaré, and Special Relativity." \emph{ArXiv}: 	arXiv:2509.09361 [physics.hist-ph], pp. 1-31. 

\end{thebibliography}
\end{document}